\documentstyle[11pt,newpasp,twoside,epsf]{article}
\def\edcomment#1{\iffalse\marginpar{\raggedright\sl#1\/}\else\relax\fi}
\reversemarginpar

\begin{document}
\title{GHASP.  A 3D Survey of Spiral and Irregular Galaxies at $H\alpha$.
\textmd{Comparison between low and high resolution rotation curves
of late-type dwarf galaxies}.}
\author{Philippe AMRAM \& Olivia GARRIDO}
\affil{Observatoire Astronomique Marseille-Provence \& Laboratoire
d'Astrophysique de Marseille, 2 Place Le Verrier, 13248 Marseille,
Cedex 4, France.}

\begin{abstract}
\noindent This survey, once completed, will provide a 3D sample of
200 nearby spiral and irregular galaxies in the $H\alpha$ line,
using a Fabry-Perot system.  The data cubes obtained for each
galaxy allow derivation of lines maps, velocity fields and higher
momentum. The goals of this survey are: (1) To constitute a 3D
local reference. (2)  To constrain the mass distribution. (3) To
constrain the kinematics and dynamics of the internal regions. A
data base will be built in order to provide the whole data to the
community.

The main result presented in this paper is that, for a sample of
19 late-type dwarfs, the \textit{HI} rotation curves corrected for
beam-smearing effects do not agree with the high resolution ones
obtained with hybrid $H\alpha$/\textit{HI} data. The beam smearing
correction done on the \textit{HI} data is on average too strong.
The rotation curves rise even more slowly when using hybrid
$H\alpha$/\textit{HI} data than with \textit{HI} data alone.

\end{abstract}

\section{Introduction}

\noindent\textbf{An Homogeneous local sample of data cube.} Up to
now, it does not exist any homogeneous sample of optical data cube
of nearby and isolated spirals with a large range of morphological
types and luminosities allowing statistical and individual
studies.  This data base will constitute an unique and homogeneous
3D sample of velocity fields and line profiles to be used as a
reference sample for:  (1) comparing nearby galaxies in various
environment (clusters, groups, pairs) with field galaxies; (2)
studying galaxies at different stages of evolution (interactions,
mergers, starbursts, ...) and  (3) analyzing galaxies presenting
anomalous motions (counter rotating populations, non keplerian
motions ...)  as well as (4) comparing high redshift galaxies for
which 3D data will be available soon with the arrival of
intermediate spectral resolution instruments on 8-meter class
telescope (e.g. GIRAFFE on the VLT).

\noindent\textbf{Mass distribution of spirals and irregulars. }
N-body simulations of cosmological evolution have now reached a
sufficient resolution to predict dark halos density profiles down
to the innermost parts of spiral galaxies. Their trend is to show
dense cuspy halos, which are not observed in most cases.
High-resolution $H\alpha$ velocity fields are complementary to
\textit{HI} velocity fields mapping the outer galactic regions but
suffering of beam smearing and of a lack of emission in the inner
regions. The GHASP sample is a sub-sample of the WHISP (Westerbork
Survey of \textit{HI} Spiral in Galaxies), lead at Groningen and
Dwingeloo, in order to map the neutral hydrogen of some 500
galaxies.   The mass distribution of the luminous and dark matter
deduced from multi-components mass models is strongly constrained
by the inner slope of the rotation curves (RCs), only correctly
drawn by 2D velocity fields once corrected for non-circular
motions. Accurate inner shape RCs should allow disentangling
cosmological scenarii (Blais-Ouellette et al. 2001).

\noindent\textbf{Simulations and theoretical interpretation of
velocity fields. } In central regions of spiral galaxies, the
presence of numerous non-axisymmetric structures like bulges
(often triaxials), bars (simple, double/nuclear or triple? bars),
circum-nuclei annulus, spiral and inner spiral, ... induces strong
velocity gradients and non-axisymmetric velocity fields.  To
understand and reproduce these structures, fine tuning models are
to be compared to the high resolution data cube. On the other
hand, with self-consistent model, it will not be possible to match
each individual velocity field with all details but it is possible
to built a grid in the space parameter (mass of the galaxies,
morphological type) in testing the good sets of initial
parameters.

In paper, section 4,  we will mainly focus on the comparison
between the RCs and the mass distribution of a sub-sample of
late-type dwarf galaxies obtained firstly in the \textit{HI}
21cm-line alone and secondly in combining $H\alpha$ and
\textit{HI}. In section 2, we will describe the galaxy sample, the
observations and the data reduction procedures. In section 3, we
will explain on several examples why one needs a 3D local
reference sample as well as for galaxies in different environments
(binaries, groups, clusters) in the local universe as for high
redshift galaxies.  In section 5, we will give some details on the
simulations. In section 6, we will conclude, summarize and provide
some issues.

\section{Galaxy sample, observations and data reductions.}

\noindent\textbf{Observations and data reduction.} Observations
began in 1998, using a scanning Fabry-Perot at the 1.93m telescope
at OHP. Similar observations have begun in March 2001, using the
same instrumentation at 1.6m telescope of Mont-M\'{e}gantic
Observatory (Qu\'{e}bec). The GHASP instrument is attached at the
Cassegrain focus of the 1.93m telescope at OHP. The original f/15
aperture ratio of the telescope is brought to f/3.9 through the
focal reducer. Interferential filters (typical FWHM 1.0 to 1.5 nm)
enable to select the $H\alpha$ line of ionized hydrogen (656.278
nm).  The $H\alpha$ line is scanned by moving the plates of the
interferometer, providing a velocity accuracy of some $km~s^{-1}$
when the signal to noise ratio is sufficient.  A Fabry-Perot
interferometer with a free spectral range of $\sim$ 380
$km~s^{-1}$ at $H\alpha$ and a effective finesse of 12 is used,
the free spectral range is scanned through 24 channels.  The total
field of view is 5.8 arcmin square and the pixel size 0.68 arcsec.
The detector is an Image Photon Counting System (GaAs IPCS). The
GHASP data are reduced with the ADHOCw software. For more detail
see Hernandez et al, this meeting; Gach et al., 2002; Garrido et
al. 2002; www-obs.cnrs-mrs.fr/interferometrie/GHASP/ghasp.html;
www-obs.cnrs-mrs.fr/interferometrie/instrumentation.html\#GaAs).

\noindent\textbf{Galaxy sample.} A sample of 200 galaxies has been
decided to be a good compromise between the necessary amounts of
data to obtain a significant sample in \textit{ad equation} with
the scientific goals and a reasonable number of observing runs for
the data acquisition on the telescope. On average, a complete
observation (including calibrations) of a galaxy can be achieved
in 2 hours.  For an eight-hours of dark time per night, 4 galaxies
can be observed per night.   At Haute-Provence Observatory (OHP,
France), the average atmospheric conditions provide an efficiency
rate of 50$\%$. With a rate of two runs of about 10-12 nights per
year,   40 galaxy per year could be obtained.  Furthermore, the
survey should go on for about 5 years. Scientifically, we expect
to cover and sample the plane (galaxy mass - galaxy morphological
type). For convenient reasons, we use the plane
luminosity-morphological type in the ranges $-16 < M_B < -23$ and
$1<T<10$. Eight galaxies per bin of 1.4 in magnitude $\times$ 2 in
morphological type are observed,  this leads to a total amount of
200 galaxies (8$\times$5$\times$5 = 200). Non-barred galaxy as
well barred galaxies are selected in each bin.  $M_B$ has been
chosen because it is a simple {\it a priori} observational
criteria, nevertheless it is not the best physical indicator of
the total mass of a galaxy , $M_R$,  $M_I$,  $M_H$, are better
ones but they are not available for a
 large range of galaxies.   Furthermore, the sample is readjusted once the
 data have been reduced in order to extract the maximum rotational velocity
 of the galaxy, which is the best indicator of the total, mass at least within
 the optical radius.  Maximum rotational velocity ranges between 50
 and 350 $km~s^{-1}$.

\section{A Local Reference sample.} An homogeneous sample of
optical data cubes of nearby and isolated spirals with a large
range of morphological types, luminosities and mass will allow
statistical and individual studies of kinematics and dynamics of
galaxies in various environments in the nearby and in the far
universe.

\noindent\textbf{The Nearby Universe.} To disentangle the
intrinsic kinematics proprieties of galaxies, including the
effects of mass, luminosity and morphological type from the
effects linked to peculiar environments, a large sample of
isolated galaxies is as much necessary as samples of peculiar
galaxies.  By peculiar environments one generally refers to the
presence of nearby galaxies or intergalactic gas, i.e. to
high-density environments (like galaxies in pairs, groups,
clusters) at different stages of evolution (like interactions,
mergers, starbursts) and sometimes presenting anomalous motions
(counter rotating populations  -star or gas-, non-axisymetric
motions and other anomalous features).  In the hierarchical
scenario, observationally confirmed by the Hubble Deep Fields
studies, a consensus is emerging that, at early epochs, all the
galaxies-in-formation were strongly interacting, at least to
collapse and form the large galaxies observed today.  In a certain
sense, any form of galaxy interactions observed today could be
considered as a nearby laboratory of the earlier universe,
especially for small galaxies, relatively metal poor, in dense
environments.  Blue Compact Galaxies (BCGs), are supposed to be
non-evolved galaxies experiencing an acceleration of their
evolution (with respect to their star formation history) at
present or recent epochs.  Except that BCGs are often relatively
isolated objects, they could be present relics of earlier epochs.
They are characterised by small to intermediate sizes, low
chemical abundances, high-star formation rate (per luminosity),
HII-region-like-emission spectra, absolute B-magnitude ranging
between -12 and -21, often surrounded by large \textit{HI} clouds.
An important issue for a better understanding of these objects is
the knowledge of the non-continuous mechanisms triggering the
present star formation. Among different hypothesis, cyclic infall
of colded gas, galaxy interactions and collapse of proto-clouds
are proposed.  From a sample of 6 BCGs + 2 star forming galaxies,
\"Ostlin \& al. (2001), Amram \& \"Ostlin (2001), proposed that
the star formation is triggering by recent merging of gas rich
objects.  If the present environment of these objects is
relatively poor, it could have been different in a recent past.  A
more complete sample of about 20 galaxies has been observed and is
presently analyzed. Nevertheless, to achieve more conclusive
results, careful calibration should be done with similar but
isolated galaxies that do not experience star formation burst.
Indeed, a massive star formation could induce strong winds
affecting the gas dynamics. Galaxies in compact group of galaxies
(CGGs) (1) are the galaxies in the more dense environment of the
nearby universe (2) experiencing high rate of galaxy-galaxy
interactions and sometimes galaxy-intergalactic gas interaction
and (3) are probably also progenitor of galaxies in formation
(tidal dwarf galaxy candidates in heavily interacting systems) in
that way, they mimic the early universe. Nevertheless, they are
different from the first galaxies because they are generally large
galaxies and are not so different from field galaxies.  It is
really a challenge to quantify in these galaxies the rate of
interaction and merging with respect to the evolution of CGGs.
From a sample of 31 groups (for a total of 104 galaxies), see
Plana et al in the present proceeding, we are going (1) to
classify the groups in different evolutionary stages of the group
(merging groups, strongly interacting groups, weakly interacting
groups, non-groups and single irregular galaxies); (2) to built
the Tully-Fisher (TF) relation CGGs and (3) to search for tidal
dwarf galaxy candidates. Here also it is absolutely obvious that
the calibration of the TF relation as well as the systematic
studies of kinematical perturbations with respect to their dense
environment need a reference sample of isolated galaxies obtained
and analysed homogeneously with the same techniques. Galaxies in
clusters are probably observed at more advanced stages of
interactions than field galaxies with respect to \textit{HI}
deficiencies, \textit{HI} and $H\alpha$ truncated disks, low star
formation rate (by the lack of fuelling gas), associated radio
sources (see e.g. Vollmer et al 2000 an example in Virgo and
Gavazzi et al. 2001 for an example in Abell 1367). In clusters, an
observational kinematical link is probably missing between the
relatively weakly disturbed spirals and the large ratio of early
type galaxies supposed to be formed from these spirals. For
cluster galaxies also, a reference sample of isolated galaxies is
dramatically needed to quantify the kinematical perturbations in
spiral -if any- and the processes of violent relation leading to
the observed high fraction of lenticular and elliptical galaxies.
The same type of arguments could be declined for shells in
early-type galaxies that are considered among the "bona fide"
signature of a past interaction event. Dynamics of warm gas
component, present in some shell galaxies combined with stellar
populations is used to constrain the accretion/interaction episode
(see Rampazzo et al in the same proceeding). Identically, binary
galaxy-galaxy interactions, for which the RCs are often non-flat
(see Isaura Fuentes-Carrera in the same proceeding) should be
interpreted through the frame of a reference sample.

\noindent\textbf{The distant universe.} Up to now, few
observations using long-slit spectroscopy at relatively low
spectral resolution with very large telescopes have revealed the
kinematics of distant galaxies. With the arrival of intermediate
spectral resolution instruments on 8-meter class telescope (e.g.
GIRAFFE on the VLT), in the next years, 3D samples of high
redshift galaxies will be studied (see Chemin et al., this
proceeding). The comparison with 3D local samples like GHASP will
be necessary to disentangle the effects of secular evolution from
the effects link to the environment like those described in the
previous section.

\section{Mass distribution of late-type dwarf galaxies.}

Dwarf galaxies are dark matter-dominated galaxies where the
stellar population make only a small contribution to the observed
RC, sometimes less important than the gaseous contribution.  It is
therefore straightforward to compare the observed RCs of these
galaxies with those derived from numerical cosmological
simulations, where the dark matter is the dominant component.

\textbf{\textit{HI} observations.} A neutral hydrogen analysis of
a sample of 60 RCs of late-type dwarf galaxies has been performed
by Swaters in his PhD thesis (1999, hereafter refereed as S99).
The S99's observations is a part of the WHISP (Westerbork survey
of \textit{HI} in SPiral galaxies) led at Westerbork
(Netherlands). Began in 1993, WHISP is a survey of the neutral
component in spiral and irregular galaxies with the Westerbork
Synthesis Radio Telescope (WSRT) at 21cm wavelength. Its aim is to
obtain maps of the distribution and velocity structure of
\textit{HI} for some 500 galaxies.  Such a uniform database will
serve as a basis for research in many areas: dark halos, effects
of environment on the structure and growth of \textit{HI} disks,
galaxy distances. More details can be found on the Web site
http://www.astro.rug.nl/$\sim$whisp/. Once computed from the
\textit{HI} velocity fields, the S99's RCs were refined
iteratively by constructing models of the observations, taking the
instrumental resolution into account. With this procedure the
author corrected for the effects of beam smearing to a large
extent.  The beam smearing plays an important role for the steeper
RCs, it depends on the inner slope of the RC and on the sampling
of the curve (Blais-Ouellette et al,1999; Blais-Ouellette et al,
2001).

\textbf{Comparison of the $H\alpha$ and \textit{HI} observations}
The data for nine galaxies over the nineteen galaxies presented in
this paper are discussed in Garrido et al. (2002, refereed
hereafter as G2002 ). The others galaxies will be discussed in
fore-coming papers. The $H\alpha$ and \textit{HI} RCs have been
combined using the following simple rule: when $H\alpha$ data are
available they have been used otherwise \textit{HI} data were
used. This means that the inner parts of the RCs have been plotted
using the $H\alpha$ velocity points while the outer parts with the
\textit{HI} data. The $H\alpha$ RCs have been shifted to the
inclinations determined for the \textit{HI} data. Best Fit Mass
models (BFMs) including three components (luminous disk, dark halo
and HI disk) have been performed with both data sets (\textit{HI}
alone and hybrid $H\alpha$+\textit{HI}). The method used in this
paper to model the mass distribution is described in Carignan $\&$
Freeman (1985), slightly generalized in Blais-Ouellette, Amram \&
Carignan (2001). The luminosity profile, in the R-band (S99), is
used to probe the mass dominant stellar component.  The luminosity
profile is transformed into a mass distribution for the stellar
disk (dash lines in Fig. 1 \& 2), assuming a constant
mass-to-light ratio $(\textit{M/L})_{R}$. For the contribution of
the gaseous component, the \textit{HI} radial profile is scaled by
1.33 to account for He (dot-dash line).  The difference between
the observed RC and the computed contribution to the curve of the
luminous (stars \& gas) component is thus the contribution of the
dark component, which can be represented by a dark spherical halo
(dotted lines). BFMs have usually been computed except when the
disk vanishes completely, in that case, a MDM is computed as
summarized in Table 2.\\

The format of this paper does not allow us to describe
individually all galaxies of the sample.  As an example, let us
briefly comment the first one.  \textbf{UGC 2023}: The \textit{HI}
RC has a solid body shape reaching almost 80 arcsec (4 kpc).  The
$H\alpha$ RC has a completely different shape as it is rather flat
up to 50 arcsec (2.5 kpc). Nevertheless, there is no $H\alpha$
emission at radius inner than 20 arcsec (1 kpc), furthermore the
kinematics of the inner regions is not so well constrained. Using
the \textit{HI} data alone, the BFM leads to a weak disk component
while the dark halo dominates at all radii! A BFM for the model
hybrid $H\alpha$/\textit{HI} RC leads to a \textit{M/L} of 0.4,
this low value is unrealistic and is due to the fact that there is
no constrain in the inner part of the RC leading to a dominant
dark halo.  The MDM for the same hybrid RC leads to a value of
\textit{M/L}=4.7 while the dark halo is, in that case, almost
negligible. In conclusion, the beam smearing effects are very
important for this galaxy even if there is no information in the
inner regions of the galaxy.
\begin{table}
\begin{tabular}{cccccccccc}
\hline
  (1)   &   (2) &    (3)    &   (4)&   (5) & (6) &  (7) &(8)&(9)&(10)
\\
\hline UGC     & NGC   &   type    &\textit{$M_B$}&D& i &$R_M$&$R_HI/H\alpha$&$R_M/h$& \textit{h} \\
\hline
2023    &       &   Im      & -16.4&   8   &   18& 2.91 & 1.43 & 2.35 & 1.24\\
2034    &       &   Im      & -17  &   11.9&   10& 4.41 & 1.88& 3.53& 1.25\\
2455    &   1156&   IB(s)m  &      &   5   &   51& 2.9& 2.06& 3.3& 0.88\\
3851    &   2366&   IB(s)m  &      &   1.3 &   40& 2.05& 1.94& 1.41& 1.45\\
4278    &       &   SB(s)dm & -19.8&   7.5 &   84& 10.78& 0.44& 4.67 & 2.31 \\
4305    &       &   Im      & -16.4&   2.1 &   45& 3.36& 2.51& 3.43 & 0.98 \\
4325    &   2552&   SA(s)m  & -17.7&   6.7 &   41& 4.68 & 1.5 &2.66 & 1.76 \\
4499    &       &   SABdm   & -16.2&   9.1 &   50& 5.95& 1.46& 4.38 & 1.36 \\
4543    &       &   Sdm     & -17  &   26  &   65& 15.1& 1.29& 4.47 & 3.38 \\
5272    &       &   Im      & -15.8&   6.9 &   65& 2.17 & 0.93& 3.39 & 0.64 \\
5414    &   3104&   IAB(s)m & -16.4&   8.2 &   45 & 4.01&0.89&2.73& 1.47 \\
5721    &   3274&   SABdm   & -17  &   7.1 &   61 & 7.76 &3.06&2.54 & 0.44\\
5829    &       &   Im      & -17.1&   8.3 &   34& 6.69 &1 &3.96& 1.69 \\
6628    &       &   Sm      & -17.5&   11.3&   25&5.75&1.09&2.16& 2.66 \\
7323    &   4242&   SAB(s)dm& -18.6&   6.9 &   40& 4.99& 1.06& 2.36& 2.11\\
7971    &   4707&   Sm      &      &   6.2 &   38 &2.26 &1.44&2.4 & 0.94\\
8490    &   5204&   SA(s)m  & -17.7&   2.6 &   50 &5.68&3.48 & 8.23& 0.69 \\
11557   &       &   SAB(s)dm& -18.4&   18.5&   37&8.07&1.04 &2.67 & 3.02\\
12060   &       &   Ibm     & -16  &   11.9&   40& 8.75&1.21&5.54 & 1.58\\
\hline
\end{tabular}\\

\noindent \textit{Description of column contents. (1) and (2) Name
of galaxy from the UGC and NGC catalogues. (3) Hubble
Morphological Type from Swaters,1999 (S99). (4) $M_B$ magnitude
from S99. (5) Distance in Mpc from S99. (6) Inclination of the
galaxy in the plane of the sky, in degrees, from S99. (7) Maximum
radius reached by the hybrid rotation curve (RC) in kpc, , using
distances given in column (5). (8) Ratio of maximum radius reached
by the \textit{HI} RC to the $H\alpha$ RC. (9) Ratio of the
maximum radius of the RC to the disk scale. (10) Disk scale length
of the optical disk in R-band, in kpc, using distances given in
column (5).}
\end{table}
\begin{table}
\begin{tabular}{c|ccc|ccc|c}
\hline
      (1)  &   (2)   &    (3)    &   (4)    &   (5)    &   (6)    &
(7)  & (8)\\
\hline
        &      &    \textit{HI} Alone    &       &       &
Hybrid\textit{HI}/$H\alpha$    &     &  \\
 UGC    &   \textit{M/L} &   $R_{0}$  &   $\rho_{0}$  &   \textit{M/L}
&   $R_{0}$  &   $\rho_{0}$& Remarks \\
\hline
2023    &   0.6 &   36    &   12  &   0.4 &   0.2 &  1101& \\
2034    &   1.7 &   9.6 &   2   &   1.3 &   3.3 &   8   &\\
2455    &   0.6 &   210.0   &   6   &   0.2 &   276.0    &   7 & \\
3851    &   2.7 &   1.7 &   20  &   0.5    &    2.3   &  20   &\\
4278    &   6.8 &   5.8 &   6  &   3.5 &   139.0   &   16  &mdm\\
4305    &   2.6 &   3 &   0   &   2.8 &   16 &   0   &\\
4325    &   5.8 &   1.8 &   32  &   8 &   9.7 &   30 & mdm\\
4499    &   0.2 &   1.5 &   52  &   1.6 &   1.9 &   32   & \\
4543    &   2.9 &   0.1 &   27  &   0.9 &   1.4    &   25  & \\
5272    &   3.2 &   8.0 &   22  &   3.1 &   55    &   26 &\\
5414    &   3.3 &   5.8 &   5  &   0.2 &   8 &   10   &mdm\\
5721    &   1.5 &   0.6 &   350 &   0.7 &   0.6 &   380 &mdm\\
5829    &   3   &   9.0 &   4   &   0.9 &   10.4 &   2.6 & \\
6628    &    1.6   & 0.1     & 191    &2.9       &0.5       &   10   &mdm\\
7323    &   1.2 &   4,0 &   17  &   2.2 &   88    &   6  & \\
7971    &   0.6 &   3 &   14  &   2.7 &   0.1 &   0   &mdm\\
8490    &   3.1 &   1.1 &   98  &   1.6 &   1.3 &   80  &\\
11557   &   0.3 &   2.5 &   24  &   0.2 &   7 &   5   &mdm\\
12060   &   3.4 &   1.3 &   47  &   0.8 &   2.3 &   45  &\\
\hline
\end{tabular}\\

\noindent \textit{Description of column contents. Best-fit mass
models has been computed except when the disk vanishes completely,
in that case, a maximum disk model is computed, see Column (8).
the dark halo is a pseudo-isothermal spheroid. (1) Name of galaxy
from the UGC catalogue. Columns (2), (3) and (4) refer to the
\textit{HI} rotation curves alone. Columns (5), (6) and (7) refer
to the hybrid $H\alpha$/\textit{HI} rotation curves. Columns (2)
and (5)refer to the disk \textit{M/L}; columns (3) and (6) to the
halo core radius in kpc; columns (4) and (7) to the halo central
density in $10^{-3}~M{\odot}~pc^{-3}$.}
\end{table}

On figure 2-right, Mass to Luminous ratio of the disk component
for the \textit{HI} RCs versus hybrid \textit{HI}/$H\alpha$ RCs
has been plotted. The \textit{HI} RC does not provide the same
value for the disk \textit{M/L} than the hybrid RC.  The huge
dispersion around the medium line (dash line) means that HI RCs
obtained after beam smearing corrections on the \textit{HI} data
do not agree with the high resolution RCs. There is no systematic
and predictable in the beam smearing correction. Nevertheless, the
disk \textit{M/L} computed from the  \textit{HI} RC is greater
than the one computed from the hybrid RC for the 2/3 of this
sample of 19 galaxies, for the other third of the sample it is the
opposite.  The beam smearing correction done on the \textit{HI}
data is on average too strong.   This  leads to an over estimation
of the disk \textit{M/L} ratio and by consequence a under
estimation of the dark halo component in \textit{HI}.  The RCs for
late type dwarf galaxies rise even more slowly when using hybrid
$H\alpha$/\textit{HI} data than with \textit{HI} data alone. The
mass of visible matter is then overestimated in \textit{HI}.

\begin{figure}[h!]
\plotone{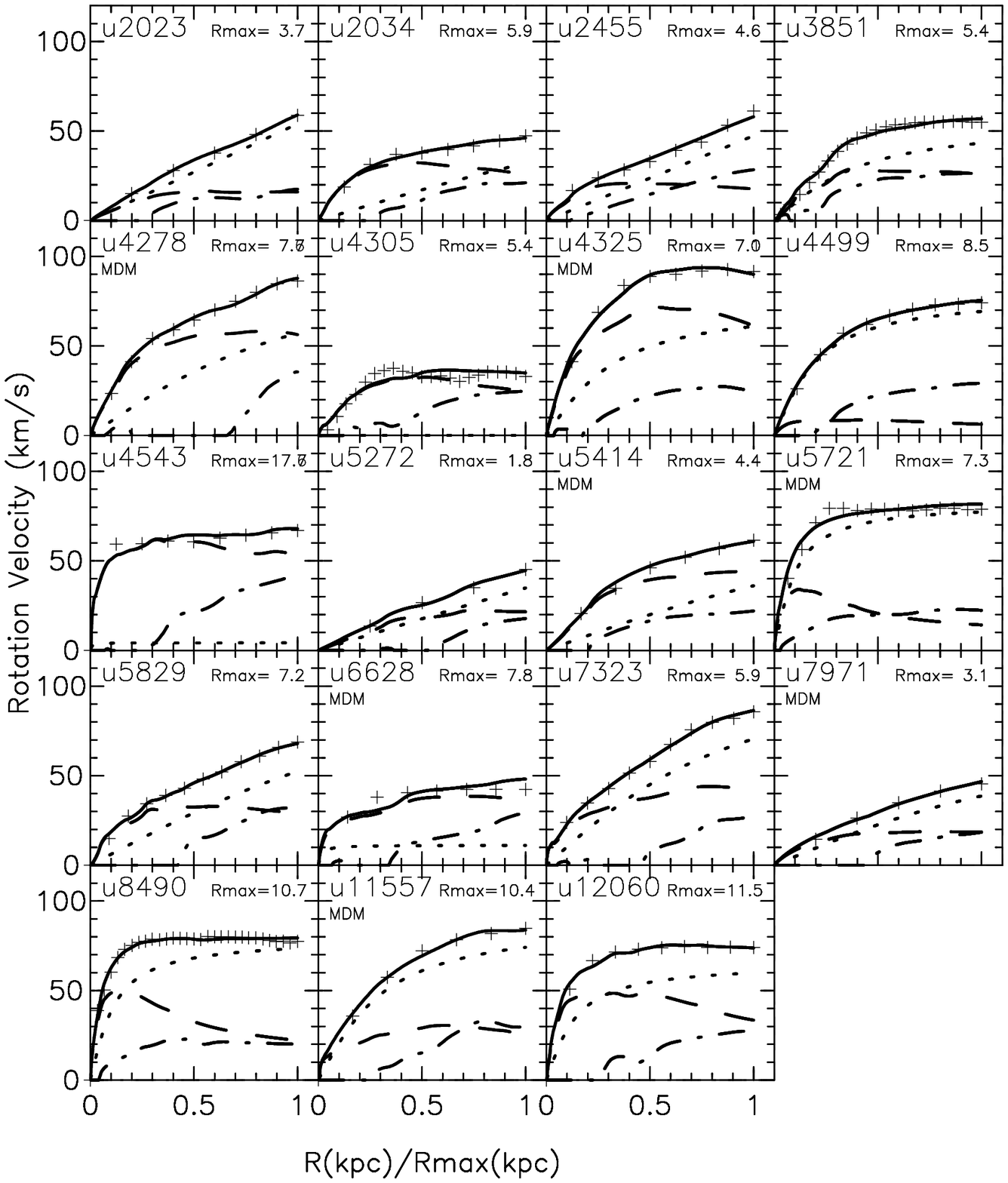} \caption{\textit{HI} Rotation curves and
mass models for a sample of 19 galaxies. Best fit models are
computed except when it leads to a disk equal to zero, it that
case a maximum disk model is computed as indicated by (MDM). The
x-axis has been scaled to the maximum radius of the rotation
curves (Rmax), Rmax are indicated for each galaxy . The symbols
"+" are the HI data; the long dashed line is the stellar disk; the
dot-dash-dot-dash is the \textit{HI} disk, the dotted line is the
pseudo-isothermal dark halo component while the full line is the
quadratic sum of the three others components.} \label{HI}
\end{figure}

\begin{figure}[h!]
\plotone{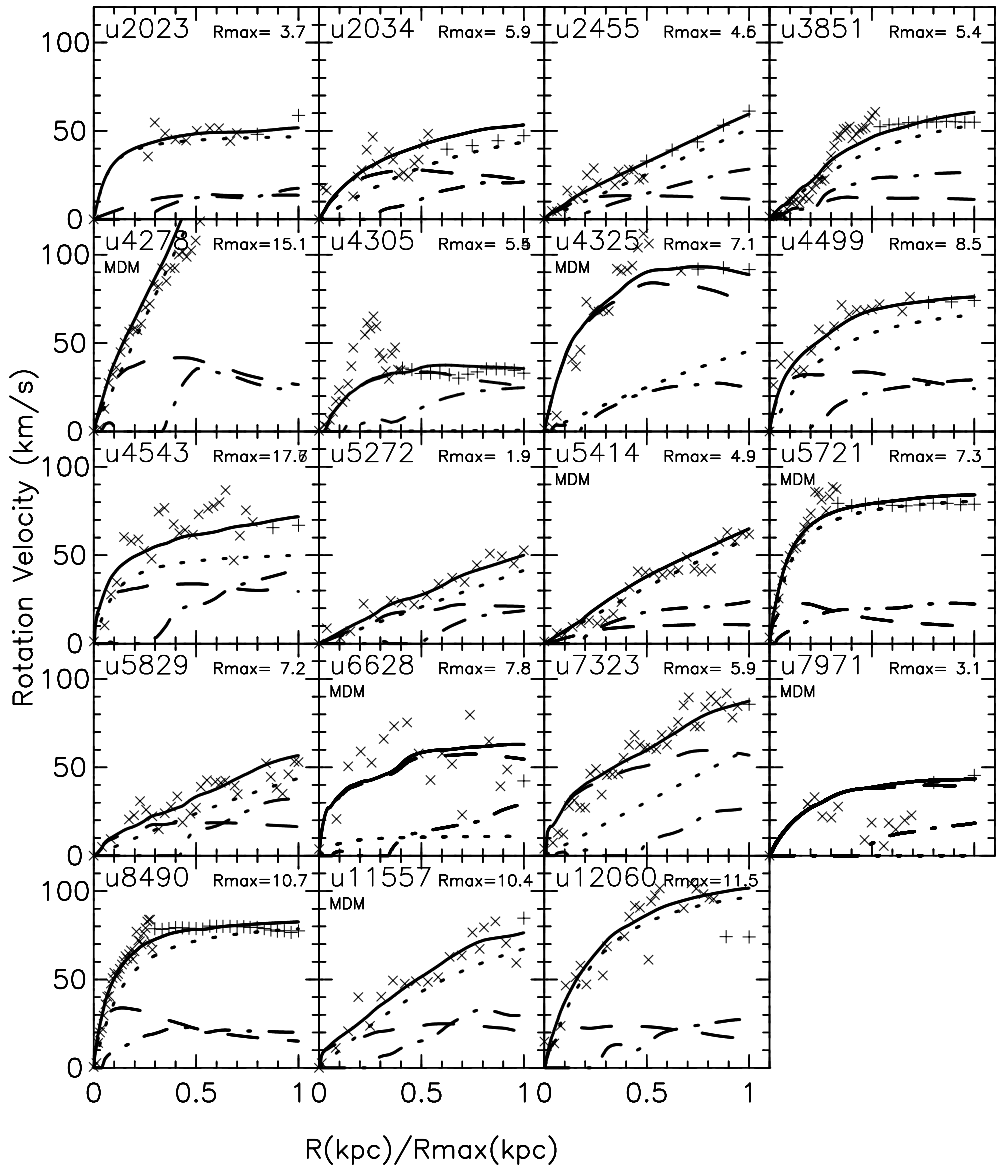} \caption{Same as Figure 1 but for the
hybrid \textit{HI}/$H\alpha$ rotation curves. The symbols
"$\times$" and "+" represent respectively the $H\alpha$ and
\textit{HI} data. Note that the rotation curve of UGC 4278 reaches
240 $km s^{-1}$ and has been truncated by the boundaries fixed for
the plot.} \label{HI+Halpha}
\end{figure}

\begin{figure}\plottwo {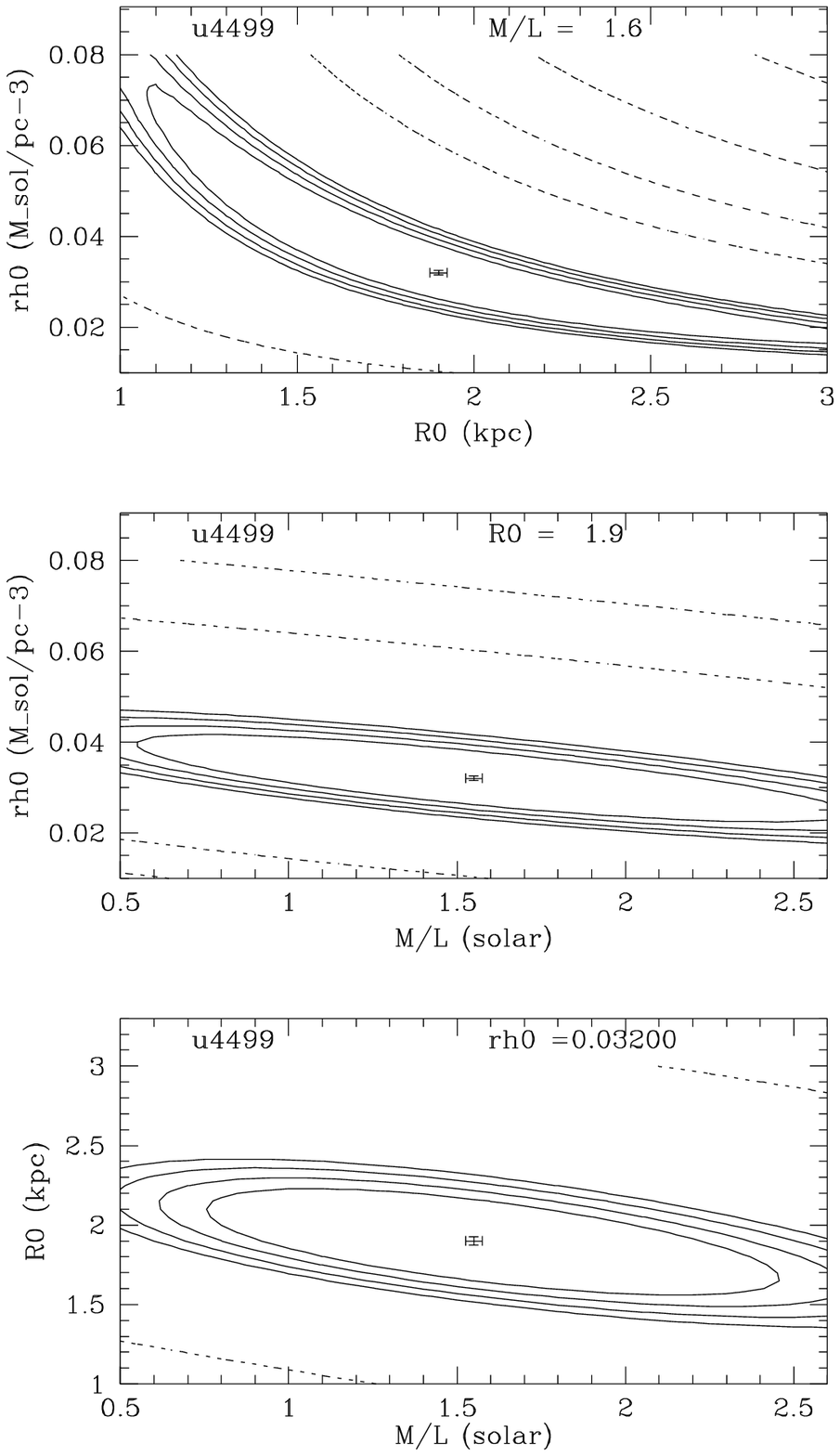}{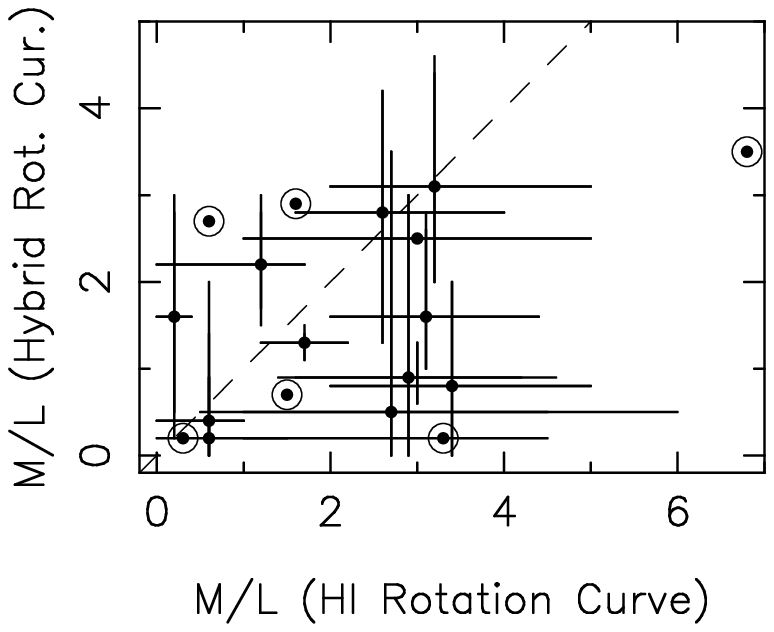}
\caption{{\bf (left):} $\chi^2$ isocontours for UGC 4499 mass
model.  The results of the best fit model is plotted here using a
pseudo-isothermal sphere. The first line around the cross
represents the one-sigma confidence level, the second line, the
two-sigma confidence level and so on. Usually, as it is presently
the case, it is clear that the \textit{M/L} parameter is the most
difficult to constrain. - {\bf (right):} Mass-to-Luminous ratio of
the disk component for the \textit{HI} rotation curves versus
hybrid \textit{HI}/$H\alpha$ rotation curves.  The error bars
represent the 3-$\sigma$ levels on the $\chi^2$ confidence
contours (like the example plotted on the left-side figure). Only
galaxies for which a best-fit model has been computed have an
error bar. For the others ones, for which a maximum disk model
(MDM) has been computed, the size of the error is represented by
the circle around the point. The dashed line represents x=y.}
\label{M/L}
\end{figure}

\section{Simulations Theoretical interpretation of velocity fields.}
Parallel to the observations, the velocity fields are modelled,
completing the works on the kinematics of the stars (Wozniak \&
Pfenniger 1997). We plan to
perform the models on two levels:\\
\textbf{(A)   }In a global way, one tries to characterize the
velocity fields according to the morphological type, mass and
environment of the galaxy. The models are generic; the results of
models are statistically compared to the observations. The codes
are adapted to the environment of studied objects. We use:
\textit{(i)} A particle-mesh N-BODY code coupled to an
hydrodynamical SPH code (Friedli \& Benz 1993, Friedli \& al.
1996), adapted to the simulations of isolated discs; \textit{(ii)}
A hierarchical N-BODY code coupled to hydrodynamical SPH code
(e.g. GADGET N-body, Springel et al, 2001) to simulate galaxies in
interaction. Previous studies never focused specifically on
kinematical issues.  With these simulations, we plan to map and
sample the plane (galaxy mass, morphological type) in same ranges
than the observations in order to match the models to the
observations. The simulations evolve freely from their start point
a couple of Gyr ago, the results of the evolution depend only of
the given initial conditions (total mass of the galaxy, amount of
star and gas, gas to star ratio, disk to halo ratio, ...). \\
\textbf{(B)   }In an individual way, one tries to model separately
each galaxy. The models are here specific. The technique is
different: the gravitational potential is determined by
photometric observations and the gas flows are simulated by the
SPH code in the gravitational potential (e.g. Sempere et al.,
1995). Although, the gas is not self-gravitating, the velocity
field remains a good approximation in the regions of weak gas
density (essentially outside nuclear regions). Finally a
theoretical and numerical special attention will be given to the
barred galaxies to disentangle the artefacts induced by the
presence of the bar in the observational RCs, when the RCs are
plotted from the 2D velocity fields without caution with respect
to the radial component along the bar.

\section{Summary, Conclusions and Issues.}
This survey will provide an unique 3D sample of about 200 nearby
spiral galaxies in the $H\alpha$ line, using a Fabry-Perot system.
The data cubes obtained for each galaxy through their moment
analysis allow us to derive crucial kinematical parameters. A data
base will be built in order to provide the whole data to the
community. The goals of this survey are:

(1) \textbf{To constitute a 3D local reference sample of
spirals/Irrs.}  It has been shown in section 3 that, with respect
to the interaction with their high density environments, no
conclusive evolutionary panel can be drawn without a solid
reference sample for cluster's galaxies, compact group galaxies,
binary galaxies, shell galaxies, star forming, ... galaxies as
well as for high redshift galaxies.

(2) \textbf{To constrain the mass distribution of spirals and
Irrs.} The present example of late type dwarf galaxies presented
here give a good idea of the impact of higher resolution rotations
curves (RCs). The main result of this analysis was to show that
the \textit{HI} RCs corrected for beam-smearing effects do not
agree with the high resolution RCs.   The beam smearing correction
done on the \textit{HI} data is on average too strong but   could
be under-estimated as well as over-estimated. The RCs for late
type dwarf galaxies rise even more slowly when using hybrid
$H\alpha$/\textit{HI} data than with \textit{HI} data alone. With
the help of the GHASP survey we will explore all the regions in
terms of galaxy mass, surface brightness and morphological types.
Other shapes for the dark haloes will be tested with the aim to
match the dark haloes computed in N-body simulations in a cosmic
evolution frame (Kravtsov et al, 1996, Navarro et al, 1996,
Burkert 1995, Burkert \& Silk 1997, Zhao, 1996). The resolution
reached in those simulations allows to predict the inner part of
halo density profiles.  These profiles could be directly compared
to the ones deduced from modelling RCs obtained in combining the
high spatial resolution of $H\alpha$ RCs and the extend of
\textit{HI} RCs. This will be done in a fore coming paper.

(3) \textbf{To constrain the kinematics and dynamics of the
internal regions.} Parallel to the observations, modelled of the
gaseous velocity fields are performed, completing the works on the
kinematics of the stars.  Simulations and theoretical
interpretation of the gaseous velocity fields in inner regions of
spiral needs an homogeneous data base.  For that purpose, on one
hand, generic models provide velocity fields characterized by
morphological type, mass and environment of the galaxy
(hierarchical N-BODY code coupled to hydrodynamical SPH code); on
the other hand, a specific model characterizes each galaxy
separately; the gravitational potential is determined by
photometric observations and the gas flows are simulated by the
SPH code in the gravitational potential.

\begin{acknowledgments}
We thank all our collaborators on the GHASP project.  In
particular M. Marcelin$^1$ for comments on the data; H.
Wozniak$^1$ for providing a simulation of NGC 3893; S.
Blais-Ouellette$^3$ and C. Carignan$^3$ for making adjustments on
their mass model programs and also C. Adami$^1$, C. Balkowski$^2$,
A. Boselli$^1$, J.Boulesteix$^1$, V. Cayatte$^2$ and L.
Chemin$^2$, J.L. Gach$^1$, O. Hernandez$^1$, L. Michel-Dansac$^1$,
H. Plana$^4$, D. Russeil$^1$ and B. Vollmer$^5$.
 $^1$Observatoire Astronomique Marseille-Provence \& Laboratoire d'Astrophysique de
 Marseille;
 $^2$GEPI, Observatoire de Paris-Meudon;
 $^3$Observatoire du Mont M\'{e}gantic \& Universit\'{e} de
 Montr\'eal;
 $^4$Observatorio Nacional, Rio de Janeiro and
 $^5$MPIfR Bonn.

\end{acknowledgments}

\begin{references}
Amram P. \& \"Ostlin G., the Messenger, 103, (March 2001)

Blais-Ouellette S., Amram P., Carignan C., 2001, AJ, 121, 4,
1952-1964.

Blais-Ouellette S., Carignan C., Amram P. \& C\^ot\'e, S., 1999,
AJ, 118, 2123.

Burkert A., 1995, ApJL, 447, L25.

Burkert A. \& Silk J., 1997, ApJ, L55.

Carignan C., Freeman K.C., 1985, AJ 294, 494.

Chemin L., Cayatte V., Flores H., Balkowski C. and Amram P.
\textit{in the present volume}.

Fuentes-Carrera I., Amram P. \& Rosado M. \textit{in the present
volume}.

Friedli, D., Wozniak, H., Ricke, M., Martinet, L., Bratschi, P.?
1996, A\&AS, 118, 461.

Friedli, D., Benz, W., 1993, A\&A, 268, 65.

Hernandez O., Gach J.L., Boulesteix J. \& Carignan C. \textit{in
the present volume}.

Gach J-L., Hernandez O., Boulesteix J., Amram P., Boissin O.,
Carignan C., Garrido O., Marcelin M., \"Oslin G., Rampazzo R.
submitted to PASP, 2002.

Garrido O., Marcelin M., Amram P. \& Boulesteix J., 2002,
submitted to A\&A.

Gavazzi G., Marcelin M., Boselli A.,  Amram P., Vilchez J.M.,
Iglesias-Paramo J. and Tarenghi M., A\&A, 2001, 377, 745.

Kravtsov A.V., Klypin A.A., Bullock J.S. et Primack, J.R. 1998,
ApJ 502, 48

Navarro J.F., Frenk C.S., White SDM, 1996, ApJ 462, 563 - 1997,
ApJ 490, 493.

\"Ostlin G., Amram P. , Masegosa J., Bergvall N., Boulesteix J.
and M\'arquez I., A\&A, 2001, 374, 800.

Plana H., Amram P., Balkowski C. \& Mendes de Oliveira C.
\textit{in the present volume}.

Rampazzo R., Amram P., Boulesteix J. \& al \textit{in the present
volume}.

Sempere M.J., Garcia-Burillo S., Combes F., Knapen J.H., 1995 A\&A
296, 45.

Springel, V., White, M., Hernquist, L., 2001, ApJ, 549, 681.

Swaters R.,A, 1999, PhD thesis, Rijksuniversiteit Groningen (S99).

Vollmer B., Marcelin M., Amram P., Balkowski C., Cayatte V.,
Garrido O., A\&A , 364, 532-542 (2000).

Wozniak H. \& Pfenniger D., 1997, A\&A 317, 14.

Zhao H., 1996, MNRAS, 278, 488.

\end{references}
\end{document}